\def\b{\begin{equation}}
\def\e{\end{equation}}

\documentstyle[eqsecnum,prd,aps]{revtex}

\begin{document}
\title{Doing it with Mirrors: Classical analogues for Black Hole radiation} 
\author{K.~Srinivasan\thanks{Electronic address:~srini@iucaa.ernet.in} 
\quad and \quad T.~Padmanabhan
\thanks{Electronic address:~paddy@iucaa.ernet.in}}
\address{IUCAA, Post Bag 4, Ganeshkhind, Pune 411 007, INDIA.}

\maketitle

\begin{abstract}
We  construct analogues for the  quantum  phenomena of black hole 
radiation in the context of {\it classical field theory}. 
Hawking radiation from a (radially) collapsing star is 
mathematically equivalent to  radiation from a mirror moving along a 
specific trajectory in Minkowski spacetime. 
We construct, in this paper, a classical analogue for 
this flat spacetime quantum phenomenon and use it 
to construct a classical analogue for black hole radiation. 

The radiation spectrum in quantum field theory has the 
power spectrum as its classical analogue.  
Monochromatic light is continually reflected off a moving 
mirror or the silvered surface of a collapsing star. 
The reflected light is fourier analysed by the observer 
and the power spectrum  is constructed. For a mirror 
moving along the standard black hole trajectory, it is 
seen that the power spectrum consists of three terms: 
(i)~a factor $(1/2)$ that is typical of the ground state 
energy of a quantum oscillator, (ii)~a Planckian 
distribution $N(\Omega)$ and---most importantly---(iii)~a term 
$\sqrt{N(N+1)}$, which is the root mean square fluctuations 
about the Planckian distribution.  
It is the appearance of the root mean square fluctuations 
which suggests  that we attribute a `thermal' nature  
to the power spectrum just like its quantum counterpart 
(which is truly thermal).  

Mirror-observer configurations like an inertial mirror 
observed in an accelerated observer's frame and an 
uniformly accelerated mirror observed in a Rindler frame 
are investigated and conditions under which  a ``thermal'' 
power spectrum is obtained are derived. 
The corresponding results in the black hole case are 
then elucidated. 
It is seen that a ``thermal'' spectrum can arise either 
due to the collapse of the star or due to the motion of 
the observer in the Schwarzchild spacetime. 
The ``temperature'' of the Planckian spectrum seen is 
therefore dependent on whether the star is in motion or 
the observer is in motion. 
In the latter case it is possible to obtain a ``temperature'' 
which is entirely independent of the mass of the star 
that is being observed.  
\end{abstract}

\reversemarginpar 

\draft


\section{Introduction and Summary}\label{sec:intro}
Hawking radiation from black holes is a quantum 
gravitational effect which is derived using the framework 
of quantum field theory in curved spacetime. 
A radially collapsing star of mass $M$ is studied close to 
its event horizon $r =2M$. 
The system is explicitly time-dependent and standard 
quantum field theoretic methods, which are applicable 
in the semi-classical limit, can be used to study it. 
Due to the collapse of the star, the in-vacuum is not the 
same as the out-vacuum and hence  the Bogoliubov coefficients 
connecting the two vacua are non-zero implying particle production. 
The radiation essentially arises due to the redshifting 
of vacuum modes  near the event horizon.  
It is shown in~\cite{unruh76} that the process by which 
Hawking radiation is derived is mathematically equivalent 
to the radiation spectrum seen due to the motion of mirrors 
in $(1+1)$--dimensional Minkowski spacetime. 
The mirror moves along a late time trajectory of the form
\b
x(t) \; = \; A - t - Be^{-Dt} \label{eqn:bhtraj}
\e
where $A$, $B$ and $D$ are arbitrary constants with $D>0$. 
We shall refer to trajectories of the above form as black 
hole trajectories since the motion of the surface of a 
collapsing star near its event horizon follows a similar trajectory. 
In this case too, the acceleration of the  mirror 
causes the redshifting of vacuum modes  giving rise to 
non-zero Bogolibov coefficients and hence particle production. 
This equivalence suggests that radiation from black holes 
can be modelled using simpler concepts from standard flat 
space quantum field 
theory~\cite{bandd82,fulling73,unruh76,carlitz87,reuter89}. 
\par
In this paper we ask whether there exist analogues 
to the above quantum phenomena in the context of 
classical field theory. 
In classical field theory,  concepts such as vacuum and 
quantum fluctuations are absent. 
To construct a viable classical model in concrete terms, 
we adopt the following procedure. 
An observer shines monochromatic light on a object 
(which can be a  mirror in flat spacetime or the silvered 
surface of a star in Schwarzchild spacetime) which is moving 
along a specified trajectory in the observer's rest frame. 
The reflected light is fourier analysed by the observer and 
the power spectrum is calculated.  We will regard the power 
spectrum as the classical analogue of the radiation spectrum 
that arises in quantum field theory in flat and curved spacetime. 
The crucial difference between the two is that the radiation 
spectrum is a function of the photon energy (expressed as 
$\hbar \Omega$ with $\Omega$ being the mean photon frequency) 
which explicitly contains $\hbar$ while the power spectrum is a 
function of the fourier transform frequency $\Omega$ alone and 
does not contain $\hbar$. As it stands, the power spectrum too 
is a well defined and measurable quantity though in the limit 
of low frequencies (like radio frequencies) or a large number 
of photons where the classical limit is applicable.  
In order to state that a particular classical power spectrum 
is the analogue of the corresponding quantum radiation spectrum,  
similarities in the {\it form} of the two spectra are considered. 
The specific terms of the power spectrum are then suitably 
interpreted based on the properties of its quantum analogue. 
The focus, in this paper, will be on the thermal radiation 
spectrum obtained in the study of collapsing stars and moving 
mirrors in quantum field theory. 
Its classical analogue is described in the following paragraph.  
\par
It is a well known result~\cite{bandd82,unruh76} that, for 
mirrors or a collapsing star travelling along a black hole 
trajectory given in Eqn.~(\ref{eqn:bhtraj}), a thermal radiation 
spectrum is obtained. 
Not only is the mean occupation number in any mode Planckian 
in form, but the fluctuations about the mean are also 
characterised by standard thermal noise.  
We look at classical scenarios in which the asymptotic 
trajectory is that given in Eqn.~(\ref{eqn:bhtraj}) and 
whose {\it power spectrum} consists of terms that motivate 
us to attribute a ``thermal'' nature\cite{klt97} to it.  
The ``thermal'' spectrum consists of three terms, none of 
which have a classical meaning. These three terms are: 
(i)~a factor $(1/2)$ that is typical of the ground state 
energy of a quantum oscillator, (ii)~a Planckian 
distribution $N(\Omega)$ and---most importantly---(iii)~a term 
$\sqrt{N(N+1)}$, which is the root mean square fluctuations 
about the Planckian distribution.  
It is due to the appearance of the root mean square fluctuations 
that the power spectrum is considered to have a `thermal' 
nature attributed to it.  
It must be emphasised again that the systems we are 
considering have no fluctuations or temperature in the 
sense of statistical physics.  Being  classical systems, 
no quantum fluctuations are present either. 
But the three terms obtained above have a natural 
interpretation in terms of notions such as thermal spectrum 
and its fluctuations. 
It was shown in Ref.~\cite{klt97} that such a ``thermal'' power 
spectrum also arises when a real, monochromatic plane wave 
of frequency $\omega$ is fourier analysed in the frame of a 
uniformly accelerating observer. Even in the limit of 
$\omega \to 0$, where the original wave tends to a constant, 
the power spectrum still remains ``thermal''.  
A constant field is the closest to what one can call 
a `classical'' vacuum.  The effect of the thermal property 
of the  power spectrum surviving even in the case of a wave 
with infinitesimal frequency can clearly be interpreted as 
the classical analogue of the quantum phenomenon in which 
a Rindler observer sees a thermal spectrum when the field 
is in the Minkowski vacuum state.  
However, in  the case of the moving mirror and collapsing 
star systems studied here, the power spectrum is found to 
be ``thermal'' strictly for non-zero incident frequencies. 
The limit of $\omega \to 0$ cannot be applied to the power 
spectum as in~\cite{klt97} and consequently one cannot envisage 
the existence of a  ``classical'' vacuum state for a black hole 
or moving mirror. 
The work presented here is a logical extension of that 
in Ref.~\cite{klt97}. For related work see 
Ref.~\cite{gerlach88,gerlach76,boyer80}
\par
We study two specific mirror-observer systems in Minkowski 
spacetime and their analogues in the Schwarzchild spacetime 
(which are star-observer systems) for which the power spectrum 
(of the reflected light) in the observer's frame of reference 
is ``thermal'' in the sense explained in the previous paragraph. 
The two mirror-observer configurations that we consider are 
(i)~an inertial mirror viewed in an accelerated observer's frame 
and (ii)~an uniformly accelerated mirror viewed in an uniformly 
accelerated observer's frame of reference. 
In case (i), the mirror always moves along a geodesic in 
Minkowski spacetime while the observer does not. 
This is analogous to the quantum system consisting of a 
Rindler observer in the Minkowski vacuum with the 
mirror as a boundary. 
Two situations relevant to a collapsing star scenario 
are considered. The first is that of an observer who is 
uniformly accelerating only for times greater than an initial 
proper time $t_i > 0$ and is inertial before that. 
Such a situation in Schwarzchild spacetime would correspond 
to a star that starts collapsing at an initial time $t_i$ 
and thereafter collapses to form a black hole. 
The second is that of an observer who is uniformly accelerating 
only between times $t_i$ and $t_f>t_i$ and is inertial otherwise. 
This system is more physically relevant since the observer 
accelerates only for a finite time interval. 
In Schwarzchild spacetime, such a situation would correspond 
to a star which starts collapsing at an initial time $t_i$ and 
stops at time $t_f$.  In other words, it does not collapse all 
the way to form a black hole. For all these cases, definite and, 
in principle, observable conditions are derived such that 
the power spectrum is ``thermal''.  
\par
Case (ii) consists of a system in which both the mirror and 
the observer are moving along non-inertial uniformly 
accelerating trajectories. The trajectories, which are both 
hyperbolae in Minkowski spacetime, are chosen so that their centers 
(the center of a hyperbola is defined as the point where the asymptotes 
of the hyperbola cross) are shifted with respect to each other. 
The motion of the mirror in the observer's frame is therefore 
explicitly time-dependent and in certain situations of relevance 
gives rise to ``thermal'' power spectra. 
The quantum analogue of such a system would possibly be that 
a Rindler observer sees particles in the vacuum of a shifted 
Rindler frame. This quantum aspect will be dealt with suitably 
in a future publication. If the centers coincide, however, the 
mirror remains stationary with respect to the observer for all 
time and the power spectrum reduces to a constant. 
The quantum analogue of this is the obvious result that a Rindler 
observer sees no particles in the vacuum of the same Rindler frame. 
The corresponding system in the Schwarzchild spacetime is that 
of a static star viewed by an observer moving away along a 
specific non-geodesic trajectory. 
This trajectory is such that the observer's frame of reference 
appears to possess a event horizon that is dependent entirely on 
the observer's motion and not on the star.  
Consequently, the ``thermal'' nature of the power spectrum  
arises solely due to the presence of this apparent horizon.   
\par
This paper is organised as follows.  In section~(\ref{sec:mirror}) 
we outline the solution for a scalar field in $(1+1)$--dimensions 
in the presence of an arbitrarily moving mirror. 
In section~(\ref{sec:accobserver}) we discuss the system of an 
inertial mirror viewed in an uniformly accelerated observer's 
frame while in section~(\ref{sec:mirrorobserver}) we study the 
system of an accelerated mirror viewed in an 
accelerated observer's frame. 
Finally in section~(\ref{sec:bh}) we study the analogues of 
the mirror configurations in the Schwarzchild spacetime. 

\section{Moving Mirrors in (1+1)--dimensions}\label{sec:mirror}
In this section we describe the purely classical problem of 
light reflected off an arbitrarily moving mirror in a 
$(1+1)$--dimensional conformal spacetime. 
We work in $(1+1)$--dimensions since, in this case, the line 
element can always be made conformally flat for any 
arbitrary spacetime metric. This conformal property 
simplifies the problem greatly since the mode functions 
of a massless scalar field in conformal co-ordinates are plane waves. 

We first briefly mention the solution to the massless 
scalar field  with the reflection boundary condition 
imposed on it (see, for instance Ref~\cite{bandd82}).
A minimally coupled massless scalar field satisfies 
the Klein-Gordon equation
\begin{equation}
\Box\Phi\equiv{\Phi^{;\mu}}_{;\mu}=0.\label{eqn:kg}
\end{equation}
In any conformally flat spacetime, the basis solutions to 
the above Klein-Gordon equation in the coordinates~$(t,x)$ 
can be taken to be plane waves labeled by the frequency $\omega$.  
Let the mirror move along an arbitrary trajectory $x_m(t)$.  
The solution to the above equation can be written in the form 
\begin{equation}
\Phi(t, x) =  e^{-i\omega(t+x)} + f((t-x)).\label{eqn:gensoln2}
\end{equation} 
This particular solution describes a situation where left 
moving plane waves, denoted by the waves $\exp(-i\omega(t+x))$, 
are incident on the mirror which acts as a boundary, and the 
resulting reflected wave, which consists of a right moving wave, 
is denoted by the as yet unknown function $f(t-x)$.  
Therefore, the solution to the right of the mirror is the 
superposition of the left moving and right moving waves 
while to the left of the mirror, the solution is 
identically zero. The reflection boundary condition is imposed 
by demanding that
\begin{equation}
\Phi(t, x_m(t)) =0 \label{eqn:gensoln3}
\end{equation}
This implies that
\begin{equation}
f(t-x_m(t)) = - e^{-i\omega(t+x_m(t))} 
\label{eqn:gensoln4}
\end{equation}
Using the definitions $u = t-x$ and $v = t+x$, the solution for 
an arbitrarily moving mirror can be written as 
\begin{equation}
\Phi(u,v) =  e^{-i\omega v} - e^{-i\omega (2\tau - u)},
\label{eqn:gensoln5}
\end{equation}
where $\tau$ is determined by solving the equation
\begin{equation}
\tau - x_m(\tau) = u 
\label{eqn:gensoln6}
\end{equation}
Since, in classical field theory, only real waves are of 
significance, the reflected light can be written in the form
\begin{equation}
\Phi_R = \cos (\omega(2\tau -u)). 
\label{eqn:gensoln7}
\end{equation}
where we have omitted the contribution from the incident waves.  
Note that, in three dimensions, or for a massive scalar field, 
one cannot use the above method to solve for the field with 
the mirror trajectory acting as the boundary. 
However, in the special case of the three dimensional wave 
vector ${\bf k}$ of the massless field being restricted to 
any one space dimension, say ${\bf k} = (\omega, 0, 0)$, then 
the above solution is obviously still valid.
\par
The above result can be rederived in an equivalent manner by 
studying the redshift of the incoming modes caused due to 
the movement of the mirror. 
Let the mirror move in an arbitrary trajectory $x_m(t)$ in 
$(1+1)$ dimensional Minkowski (or conformal) spacetime.  
Consider an observer positioned at the origin and shining a 
beam of light with the frequency $\omega$ at the mirror. 
To be specific, let the mirror be situated to the {\it left} 
of the observer. 
The incident beam of light is represented as  $(\omega, k)$ 
with the wave vector $k = -\omega$.  
Let the light beam reach the mirror at the time $t_0$. 
At this time, the mirror has a velocity $v_m(t_0)$ and 
therefore, in a Lorentz frame moving with the velocity $v_m(t_0)$ 
in which the mirror is instantaneously at rest, the incoming 
light beam appears to be doppler shifted with a new frequency 
$\omega'$ and wave vector $k'$ given by the relations
\begin{equation}
\omega' \; = \; \omega \sqrt{ 1 + v_m(t_0) \over 1 - v_m(t_0)} 
\qquad ; \qquad k' \; = \; -\omega \sqrt{ 1 + v_m(t_0) 
\over 1 - v_m(t_0)} \label{eqn:ref1}
\end{equation}
After reflection, the reflected wave has the same frequency 
$\omega'$ in the instantaneous Lorentz frame, but the wave 
vector changes sign to $-k'$. Therefore, the quantities 
$(\omega_R, k_R)$, which refer to the reflected light in 
the observer's frame of reference, are
\begin{equation}
\omega_R \; = \; \omega' \sqrt{ 1+v_m(t_0) \over 1 - v_m(t_0)} 
\qquad ; \qquad k_R \; = \; \omega' \sqrt{ 1+v_m(t_0) 
\over 1 - v_m(t_0)} \label{eqn:ref2}
\end{equation}
where the corresponding {\it inverse} Lorentz transformations 
have been used to transform back to the observer's frame.  
Substituting for $\omega'$ from Eqn.~(\ref{eqn:ref1}) into 
Eqn.~(\ref{eqn:ref2}), one obtains,
\begin{equation}
\omega_R = \omega \left( {1+v_m(t_0) \over 1 - v_m(t_0)}\right) 
\qquad ; \qquad  k_R \; = \; \omega \left( {1+v_m(t_0) 
\over 1 - v_m(t_0)}\right) \label{eqn:ref3}
\end{equation}
Let this reflected light reach the observer at time $t>t_0$. 
The phase of the reflected wave $\theta(t)$ reaching the 
observer as a function of $t$ is given by
\begin{equation}
\theta(t) \; = \; \int dt \,\omega_R(t_0(t)) \; = \; \omega 
\int dt \left( {1+v_m(t_0) \over 1 - v_m(t_0)}\right) 
\label{eqn:ref4}
\end{equation}
The relationship between $t$ and $t_0$ is quite obviously 
\b
t \; - \; t_0 \; = \; -x_m(t_0) \label{eqn:ref5} 
\e
where we have taken into account the fact that the mirror is 
situated to the left of the observer and $x_m(t_0)$ is the 
distance between the mirror and observer at time $t_0$. 
The above equation is exactly the same equation as in 
Eqn.~(\ref{eqn:gensoln5}) with $u$ replaced by $t$ (the observer 
is seated at the origin and so $u = t-x \equiv t$) and $\tau$ 
replaced by $t_0$.  If the observer were situated at some 
other position $x$, then it is easy to see that the above 
equation still holds with $t$ replaced by $t-x$ since the 
wave reaching the observer is a right moving wave. 
From Eqn.~(\ref{eqn:ref5}) we have,
\begin{equation}
{dt_0 \over dt} \; = \; {1 \over 1 - v_m(t_0)} 
\label{eqn:ref6}
\e
Using this in Eqn.~(\ref{eqn:ref4}), one has
\b
\theta(t) \; = \; \omega \left[ 2 \int \! {dt \over 1-v_m(t_0)} 
\; - \; t \right] \; = \; \omega (2t_0 - t) 
\label{eqn:ref7}
\e
with $t_0$ being given in terms of $t$ by solving 
Eqn.~(\ref{eqn:ref5}). Thus, the phase of the reflected wave 
given in the above equation matches that  in 
Eqn.~(\ref{eqn:gensoln4}) with $u$ replaced by $t$ and $\tau$ 
replaced by $t_0$.   \par
With the basic formalism outlined in this section, we move on 
to describing scenarios where the mirror trajectory assumes 
the trajectory given in Eqn.~(\ref{eqn:bhtraj}) which gives 
a power spectrum that is ``thermal'' in form.

\section{Inertial mirror viewed in an Accelerated observer's frame}
\label{sec:accobserver}
In this section, we study a system consisting of a mirror 
at rest in Minkowski spacetime viewed in an uniformly 
accelerated observer's frame of reference. 
The mirror moves along a geodesic while the motion of 
the observer is non-geodesic. Consider an observer who is 
accelerating uniformly in Minkowski spacetime with 
an acceleration $g>0$. 
Let the co-ordinates of the observer's frame of reference be 
denoted by the conformal co-ordinates $(t,x)$.  
The transformations~\cite{rindler66} between the Minkowski 
co-ordinates $(T,X)$ and the Rindler co-ordinates $(t,x)$ are 
\begin{equation} 
T = T_o + g^{-1}e^{gx}\,\sinh(gt)\quad ; \quad 
X = X_o + g^{-1}e^{gx}\,\cosh(gt)
\label{eqn:rindcoords} 
\end{equation}
where $X_o$ and $T_o$ are arbitrary constants. 
The trajectory of an observer at rest in the accelerated 
frame is a hyperbola whose center (the center of a 
hyperbola is defined as the point at which the asymptotes 
of the hyperbola cross) is shifted from the Minkowski 
spacetime origin by the vector $(T_o, X_o)$. 
Let the mirror be at rest at a spatial distance $X_m$ from 
the origin of co-ordinates. 
The trajectory of the  mirror as viewed in the Rindler 
frame is 
\begin{equation}
X_m \; = \;  X_o + g^{-1}e^{gx_m(t)}\,\cosh(gt) 
\label{eqn:exactt2}
\end{equation}
which implies
\begin{equation}
gx_m(t) = \ln\left[g(X_m-X_o)\right] - 
\ln\left[\cosh(gt)\right], 
\label{eqn:exactt3}
\end{equation}
Note that $X_m$ has to be chosen so that $X_m - X_o > 0$. 
This is necessary because otherwise the reflected light 
will never reach the Rindler observer. (If $X_m = X_o$, 
the mirror is always at the event horizon $x_m = -\infty$ 
for all time and hence not of interest.) Since only the 
combination $(X_m - X_o)$ appears, we can set $X_o = 0$ 
without any loss of generality. Therefore, as observed 
in the Rindler frame, the mirror advances from the event 
horizon (for times $t \to -\infty$) to a point $x_m = g^{-1} 
\ln(gX_m)$ (at $t = 0$) and recedes back into the horizon 
(for times $t\to \infty$).  
Let the observer be seated at a distance $x_0$  (with $u=t-x_0$). 
To avoid a collision with the mirror one must have 
$x_0 > \;  g^{-1} \ln(gX_m)$. The Rindler observer 
continuously shines monochromatic light of frequency 
$\omega$ on the moving mirror and receives the reflected light.  
Solving Eqn.~(\ref{eqn:gensoln6}), the reflected wave 
(see Eqn.~(\ref{eqn:gensoln7})) is 
\begin{equation}
\Phi_R \; = \; \cos\left[ \omega g^{-1}\ln\left(2g X_m\,e^{gu} - 
1\right) \, - \omega u \right] 
\label{eqn:exactref1}
\end{equation}
The exact fourier transform of this real wave with respect to 
$t$ along the entire trajectory of the mirror yields a fourier 
amplitude given in terms of Beta functions which is not of 
interest here. In a collapse scenario, such a model would  
correspond to a star expanding from the event horizon to a 
finite radius and then collapsing back again into the horizon.  
This is clearly not a physically relevant situation. 
Instead, we focus on the mirror trajectory at late times 
$t > (1/g)$ when it is receding into the event horizon. 
Notice that this late time condition is similar to the 
condition $t > ({\rm surface \; gravity})^{-1} = (4M)^{-1}$ 
for Hawking radiation to arise for a collapsing star. 
The role of the ``surface gravity'' is played by 
$g$ in this case.  So, for late times, one has, 
\begin{equation}
gx_m(t) \; = \; \ln(2gX_m) - gt - e^{-2gt} + {\cal O}
\left( e^{-4gt}\right) \label{eqn:latetimes}
\end{equation} 
where ${\cal O}$ is the order symbol which indicates 
the order of the terms being neglected. In an expression 
of the form $Q = R + {\cal O}(x^\alpha)$, the term 
${\cal O}(x^\alpha)$ indicates that $Q = R$ to order 
$x^\alpha$.  All powers of $x$ greater than or equal 
to $\alpha$ are ignored. 
The reflected wave reduces to,
\b
\Phi_R \; = \; \cos\left[ \omega A - \omega B e^{-gt} + 
{\cal O}\left(e^{-2gt}\right) \right] 
\label{eqn:latetimes1}
\e
where
\b
A \; = \; g^{-1}\, \ln(2gX_m) \qquad ; \qquad 
B = {1 \over 2g^2 X_m}\, e^{gx_0} 
\label{eqn:latetimes2}
\e     
When taking the fourier transform of the reflected wave, 
we integrate, not from $(-\infty, \infty)$, but from 
$(t_i, \infty)$ where $t_i \sim (1/g)$.(In calculating 
the order in Eqns.~(\ref{eqn:latetimes1}, \ref{eqn:latetimes}) 
the quantity $x_0$ in $u=t-x_0$ has been assumed to be either 
negative or small compared to $(1/g)$ if positive. 
If $x_0 \sim (1/g)$ say, the reflected wave can assume 
the form given in Eqn.~(\ref{eqn:latetimes1}) only if 
the order of the terms neglected is ${\cal O}(e^{-2gu})$. 
This in turn implies that the late time trajectory has to 
hold for $t> x_0 + (1/g)$ and not just for $t>(1/g)$.) 
It is important to note that we have assumed that the 
trajectory for times $t < t_i$ is inertial. 
Consequently the contribution to the reflected wave from 
times $t< t_i$ can be safely ignored. In a realistic 
collapse scenario, this model would correspond to a stationary 
star that starts collapsing at a finite time and which 
collapses all the way to form a black hole and thus 
is physically feasible.  
To compute the  fourier transform, it is convenient to 
decompose the reflected wave as follows:
\b
\Phi_R = \psi_R + \psi_R^* 
\label{eqn:decompos}
\e
where 
\b
\psi_R \; = \; {1 \over 2}\, e^{i\omega A}\, 
e^{-i\omega B e^{-gt}} 
\label{eqn:latetimes3}
\e
with a corresponding expression for $\psi_R^*$. 
The fourier transform of $\psi_R$ is given by the expression,
\b
\tilde{\psi}_R\; = \; \int_{t_i}^{\infty} \! dt e^{-i\Omega t} 
\psi_R \; = \; {1 \over 2g}\, e^{i\omega A}\, 
(\omega B)^{-i\Omega/g} \int_{0}^{\omega B e^{-gt_i}} 
\! {dy \over y} \, y^{i\Omega/g} \, e^{-iy} 
\label{eqn:latetimes4}
\e
where we have made the change of variable 
$y = \omega B e^{-gt}$.  
Evaluating $\tilde{\psi}_R$ by appropriately rotating in 
the complex plane, one obtains,
\b
\tilde{\psi}_R (\Omega) \; = \; {1 \over 2g}\, e^{i\omega A}\, 
(\omega B)^{-i\Omega/g} \, e^{\pi \Omega/2g}
\left\{ \Gamma[i\Omega/g] \; - \;  {\cal O}
\left({e^{-\omega B e^{-gt_i}} \over \omega B e^{-gt_i} }\right) 
\right\} 
\label{eqn:latetimes5}
\e
where $\omega$ must be chosen appropriately so that the 
above relation is correct to  
${\cal O}(e^{-\omega B e^{-gt_i}}/\omega B e^{-gt_i})$. 
This implies that $\omega$ cannot be arbitrarily small. 
It must at the very least satisfy the inequality 
$\omega > B e^{-gt_i}$. 
Evaluating the fourier transform of $\psi_R^*$ in a similar 
manner and adding it to the expression for $\tilde{\psi}_R$, 
the fourier transform of $\Phi_R$ is given by,
\b
\tilde{\Phi}_R (\Omega) \; = \; {1 \over 2g}\, 
(\omega B)^{-i\Omega/g} \, \Gamma[i\Omega/g]\, 
\left( e^{i\omega A}\, e^{\pi \Omega/2g} \; + \; 
e^{-i\omega A}\, e^{-\pi \Omega/2g} \right)
\label{eqn:latetimes6}
\e 
The power spectrum per unit logarithmic frequency 
interval~\cite{klt97} is,
\begin{equation}
{\cal P}(\Omega)\equiv \Omega\; {\vert{\tilde \Phi}
(\Omega)\vert}^2
= \left({\pi\over g}\right)\;
\; \left\lbrace{1 \over 2} + N
+ \sqrt{N(N+1)}\; \cos(2\beta)\right\rbrace 
\label{eqn:ps}
\end{equation}
where $N$ is defined to be 
\begin{equation}
N(\Omega)=\left({1 \over 
{\exp\left(\Omega/\Omega_0\right)-1}}\right).
\label{eqn:planck}
\end{equation}
and 
\begin{equation}
\Omega_0 = g/2\pi \qquad ; \qquad \beta = \omega A 
= \omega g^{-1}\ln(2 gX_m) 
\label{eqn:latetimes7}
\end{equation}
where we have substituted the expressions for $A$ 
and $B$ from Eqn.~(\ref{eqn:latetimes2}). 
This power spectrum, as explained in 
section~(\ref{sec:intro}), has the appearance of a 
``thermal'' spectrum.  
It can be therefore regarded as a fit classical 
analogue of its quantum counterpart, the radiation spectrum. 
(For a detailed discussion see Ref.~\cite{klt97}.) 
The incident frequency $\omega$ must at least satisfy 
an inequality of the form 
\begin{equation}
\omega > {2g^2X_m \over c^3} e^{g(t_i - x_0)/c} 
\label{eqn:ftrel}
\end{equation} 
where the correct factors of $c$ have been put in, 
for the power spectrum to have the form above.   
\par
Note that since the trajectory~(\ref{eqn:exactt3}) is symmetric 
in the time co-ordinate, considering the {\it early} 
time behaviour of the mirror trajectory gives the same 
power spectum per unit logarithmic frequency interval as 
the late time behaviour did. This, however, is not of 
physical relevance since such a model would correspond 
to a star expanding outwards from the event horizon 
to a finite radius.   
\par
We now analyse a situation in which the mirror has the 
trajectory given in Eqn.~(\ref{eqn:latetimes}) for a 
given time interval and ask under what conditions 
would a spectrum of the form in Eqn.~(\ref{eqn:ps}) 
be obtained.  
To make things more specific, consider a mirror 
trajectory of the form
\begin{equation}
gx_m(t) \; = \; \ln(2gX_m) - gt - e^{-2gt} \qquad 
\qquad t_i \; < \; t\; < \; t_f 
\label{eqn:latettimes}
\end{equation}
for  $t_i < t < t_f$ and inertial otherwise. 
In Minkowski space, this corresponds to the observer 
accelerating uniformly only between the times $t_i$ and 
$t_f$ (and moving along an inertial trajectory for $t< t_i$ 
and $t > t_f$) with the mirror placed as before at the 
spatial distance $X_m$. The mirror trajectory in 
Eqn.~(\ref{eqn:latettimes}) is a good approximation to the 
exact trajectory in Eqn.~(\ref{eqn:exactt3}) if $t_i \sim (1/g)$. 
The reflected waveform is of the same form as in  
Eqn.~(\ref{eqn:latetimes1}). Decomposing it into the 
form given in Eqn.~(\ref{eqn:decompos}), the fourier 
transform of $\psi_R$ is now given by
\begin{equation}
\tilde{\psi}_R(\Omega) \; = \; \int_{t_i}^{t_f} \! dt 
e^{-i\Omega t} \psi_R \; = \; {1 \over 2g}\, e^{i\omega A}\, 
(\omega B)^{-i\Omega/g} 
\int_{\omega B e^{-gt_{\! f}}}^{\omega B e^{-gt_{\! i}}} \! 
{dy \over y} \, y^{i\Omega/g} \, e^{-iy}  
\label{eqn:ftt1}
\end{equation}
where the usual change of variable $y = \omega B e^{-gt}$ 
has been made. 
Setting $\omega B e^{-gt_{\! f}} = \epsilon$, the above 
integral can be evaluated to give,
\begin{eqnarray}
\tilde{\psi}_R(\Omega) \; &=& \; {1 \over 2g}\, e^{i\omega A}\, 
(\omega B)^{-i\Omega/g} \left\{ \; e^{\pi \Omega/2g}\, 
\left[\Gamma[i\Omega/g] \; - \;   
{\cal O}\left({e^{-\omega B e^{-gt_i}} 
\over \omega B e^{-gt_i} }\right) \right] \right. 
\nonumber \\
 && \qquad \qquad \qquad \qquad \qquad \qquad \qquad 
\qquad \quad - \; \left. {\epsilon^{i\Omega/g} 
\over {i\Omega/g}} \: + \; {\cal O}(\epsilon) \quad \right\} 
\label{eqn:ftt2}
\end{eqnarray}
Evaluating $\tilde{\psi}_R^*$ in a similar manner and 
neglecting the correction terms, the expression for 
$\tilde{\Phi}_R(\Omega)$ can be written as,
\b
\tilde{\Phi}_R (\Omega) \; = \; {1 \over 2g}\, 
(\omega B)^{-i\Omega/g} \, \left( e^{i\omega A}\, 
e^{\pi \Omega/2g} \; + \; e^{-i\omega A}\, 
e^{-\pi \Omega/2g} \right) \Gamma[i\Omega/g] \; + \; 
{i \over \Omega} \, \cos(\omega A) \, e^{-i\Omega t_f} 
\label{eqn:ftt3}
\e
where we have substituted for $\epsilon$ in terms of $t_f$. 
The power spectrum per unit logarithmic frequency interval 
is easily calculated to be 
\begin{eqnarray}
{\cal P}(\Omega)\equiv \Omega\; 
{\vert{\tilde \Phi}_R(\Omega)\vert}^2 &=& 
\left({\pi\over g}\right)\; \left\lbrace{1 \over 2} + 
N + \sqrt{N(N+1)}\, \cos(2\beta)\right\rbrace \; + 
\; {1 \over \Omega} \, \cos^2(\omega A) \nonumber \\
&+& \sqrt{\pi \over 2g \Omega} \, \cos(\omega A) 
\left\lbrace \cos(\omega A) \, \sin(\alpha)\, 
\sqrt{1 + 2N} \; + \; \sin(\omega A) \cos(\alpha) 
{1 \over \sqrt{1 + 2N}} \right\rbrace 
\label{eqn:ftt4}
\end{eqnarray}
where $N$ is defined as in Eqn.~(\ref{eqn:planck}) 
and the quantity $\alpha$ is given by
\b
\alpha \; = \; \Omega \left(t_f - g^{-1}\ln (\omega B) 
\right) \; + \; {\rm arg} \left(\Gamma[i\Omega/g] 
\right) \label{eqn:ftt5}
\e
To neglect the correction terms in Eqn.~(\ref{eqn:ftt2}), 
$\omega$ must at least satisfy the inequality given in 
Eqn.~(\ref{eqn:ftrel}) while  $t_f$ should be chosen so 
that  $t_f - t_i > (1/g)$. If $t_f$ is chosen to be large 
enough, $\alpha \to \infty$ and so the second term in 
curly braces can be set to zero. But there is still a 
term proportional to $1/\Omega$ which diverges for 
small $\Omega$. However, notice that when $\Omega > g$, 
$N(\Omega) \simeq e^{-\Omega/\Omega_0}$.  
When this is satisfied, both the extra terms can be neglected 
and one obtains the usual thermal spectrum. Summing up, 
one sees that for a mirror moving along the 
trajectory~(\ref{eqn:latettimes}) for times $t_i < t < t_f$, 
the power spectrum matches that given in Eqn.~(\ref{eqn:ps}) 
for $\Omega > g$ when the conditions $t_i\sim (1/g)$, 
$t_f-t_i > (1/g)$ and the inequality in Eqn.~(\ref{eqn:ftrel}) 
are satisfied.

\section{Accelerated mirror viewed in an Accelerated observer's frame} 
\label{sec:mirrorobserver}
We next consider the system of an uniformly accelerated 
mirror viewed in an uniformly accelerated observer's frame. 
The trajectories of both the observer and the mirror are 
hyperbolae in Minkowski spacetime. However, the 
centers\footnote{The center of a hyperbola is the point 
at which its asymptotes cross. For the usual Rindler observer, 
the center is just the Minkowski spacetime origin.} of both 
these hyperbolae are shifted.  Such a situation is an explicitly 
time-dependent one with the mirror moving from in from the  
event horizon of the observer and receding back. 
We will concentrate as usual only on the receding part 
of the mirror's trajectory. 
\par
The transformations between the Minkowski co-ordinates $(T,X)$ 
and the Rindler co-ordinates $(t,x)$ of the observer's frame 
are given as usual by
\begin{equation} 
T = T_o + g^{-1}e^{gx}\,\sinh(gt) \quad ; \quad 
X = X_o + g^{-1}e^{gx}\,\cosh(gt) 
\label{eqn:rindcoords1} 
\end{equation} 
where $g$ is the acceleration of the observer and $T_o$ and  
$X_o$ are arbitrary constants. We assume $X_o>0$ here and 
discuss the case $X_o < 0$ later. Thus, the center of the 
hyperbolic trajectory of a observer seated at a distance 
$x_0$ is at $(T_o,X_o)$. The trajectory of the mirror in 
Minkowski space is chosen to be the hyperbola 
\begin{equation}
X^2 - T^2 = g_m^{-2} \label{eqn:amo2}
\end{equation}
where $g_m$ is the acceleration of the mirror and whose center 
is at the origin. This can always be done without any loss 
of generality since only the relative motion between the 
mirror and the observer is of consequence. 
We assume that the mirror is confined  to the right 
Rindler sector with $X>0$ always. The trajectory of the 
mirror in the observer's frame is obtained by substituting 
for $T$ and $X$ from Eqn.~(\ref{eqn:rindcoords1}) 
into Eqn.~(\ref{eqn:amo2}). Doing this one obtains 
the following quadratic equation in the variable $e^{gx_m(t)}$.
\begin{equation}
g^{-2} e^{2g x_m} + 2 g^{-1} e^{g x_m}\left[ X_o 
\cosh(g t) - T_o \sinh(g t) \right] = g_m^{-2} - 
( X_o^2 - T_o^2)  
\label{eqn:amo3}
\end{equation}
We now set $T_o = 0$ for simplicity. The case $T_o \neq 0$ 
offers no new possibilities and will not be 
considered further. The above equation is easily 
solved to give
\b
e^{gx_m(t)} \; = \; -gX_o\cosh(gt) \; \pm \; 
\sqrt{g^2X_o^2\cosh^2(gt) + g^2(g_m^{-2} - X_o^2)} 
\label{eqn:amo4}
\e
Notice that only the positive sign leads to real $x_m(t)$ 
since $X_o > 0$. If $X_o = 0$, then the solution reduces 
very simply to $x_m(t) = g^{-1}\ln(g/g_m)$. 
The  mirror remains  at a constant distance in the 
observer's frame for all time. In Minkowski space, 
the centers of the hyperbolic trajectories of the observer 
and mirror coincide and one gets the trivial result of a 
constant power spectrum. The quantum analogue of this result 
merely states that a Rindler observer sees no particles 
in the Rindler vacuum.  This can be made more precise by 
constructing the Wightman function in the Rindler frame 
(which is identical to that in Minkowski space since the 
Rindler frame is conformal to it) for a trajectory that 
is inertial in the observer's frame. 
The detector response is determined by fourier transforming 
the Wightman function.  The expected answer is zero.  
Hence, this classical result does have a corresponding 
quantum analogue. 
\par
For a non-trivial time-dependent solution for $x_m(t)$ 
to exist, one must have $(g_m^{-2} - X_o^2) > 0$. 
(The case $(g_m^{-2} - X_o^2) = 0$ just corresponds to a 
situation in which the mirror is at the event horizon of 
the observer's frame for all times since $x_m(t) = -\infty$.) 
Since a collision between the observer sitting at the point 
$x_0$ and the mirror is unwanted, $x_0$ should satisfy 
the inequality,
\b
x_0 \; > \; g^{-1} \ln\left( g[g_m^{-1} - X_o]\right) 
\label{eqn:amo4a}
\e
For late times $t>0$ such that 
\b
\cosh(gt) \gg \sqrt{g_m^{-2} - X_o^2 \over X_o^2}, 
\label{eqn:amo4b}
\e
we have,
\b
e^{gx_m(t)} \; = \; -gX_o\cosh(gt) \; + \; gX_o\cosh(gt) 
\sqrt{1 + {g_m^{-2} - X_o^2 \over X_o^2\cosh^2(gt)} } \; 
\approx \; {g(g_m^{-2} - X_o^2) \over 2X_o\cosh(gt)} 
\label{eqn:amo5}
\e
which implies that
\b
g x_m(t) \; = \; \ln\left[ {g(g_m^{-2} - X_o^2) \over 2X_o} 
\right] \; - \; \ln[\cosh(gt)] \label{eqn:amo6}
\e 
Therefore, one sees that the trajectory of the mirror as 
seen by the observer resembles that of a inertial mirror 
viewed in an uniformly accelerated observer's frame which 
was treated in the last section. Such a trajectory has exactly 
the same form as that in Eqn.~(\ref{eqn:exactt3}). 
Hence, the late time power spectrum per unit logarithmic 
frequency interval has the same form as that in 
Eqn.~(\ref{eqn:ps}) but with $\beta$ given by
\b
\beta = \omega g^{-1}\ln({g(g_m^{-2} - X_o^2) \over X_o}) 
\label{eqn:amo7}
\e  
The same condition on $\omega$ in Eqn.~(\ref{eqn:ftrel}) 
must hold for the spectrum to be thermal. 
Knowing the classical result we can attempt to predict the 
analogous quantum result. The classical result appears 
to indicate that the vacua of two Rindler frames whose 
centers in Minkowski spacetime are shifted are inequivalent. 
This means that the  Bogolibov coefficients between the 
two vacua must be non-zero implying particle production. 
Since the classical power spectrum is ``thermal'', the 
quantum radiation spectrum can also be expected to be 
thermal in nature. An observer present in one Rindler 
frame therefore sees a thermal distribution of photons 
when the scalar field is in the vacuum of the shifted frame. 
This result is characteristic of the Minkowski-Rindler system 
and the connection does not appear to be an obvious one.  
We will explore this quantum analogue more fully in 
a future publication.  
\par
We now briefly discuss the case when $X_o < 0$.  
Setting $X_o = -|X_o|$ in Eqn.~(\ref{eqn:amo4}), 
one finds that 
\b
e^{gx_m(t)} \; = \; g|X_o|\cosh(gt) \; \pm \; 
\sqrt{g^2|X_o|^2\cosh^2(gt) + g^2(g_m^{-2} - |X_o|^2)} 
\label{eqn:amo8}
\e
The simplest case above is when $g_m^{-2} - |X_o|^2 = 0$.  
The only non-trivial  solution is (the trivial solution is 
$x_m(t) = -\infty$ which is uninteresting since it implies 
that the mirror remains at the event horizon for all time)
\b
x_m(t) \; = \; g^{-1}\ln\left({2g \over g_m}\right) \; + 
\; \ln[\cosh(t)] \label{eqn:amo9}
\e  
The crucial difference between the above solution and 
that in Eqn.~(\ref{eqn:amo6}) is the presence of a positive 
sign before the $\ln[\cosh(t)]$ term. 
Note that the mirror is always to the {\it right} 
of the observer. To avoid a collision between the mirror 
and the observer, the observer has to be stationed at 
points $x_0 < g^{-1}\ln(2g/g_m)$.  
Modifying the formalism developed in section~(\ref{sec:mirror}) 
appropriately (replacing $u=t-x$ by $v=t+x$) to take into 
account this case, one finds that the late time power spectrum 
(for $t>1/g$ as usual and making the usual approximations 
made in section~(\ref{sec:accobserver})) is exactly the same 
as that in Eqn.~(\ref{eqn:ps}) with 
$\beta = \omega g^{-1}\ln(g/g_m)$.  
This shows that as long as the trajectory is of the 
form in Eqn.~(\ref{eqn:bhtraj}), one obtains a thermal power 
spectrum regardless of the presence of a event horizon or not. 
The above solution naturally cannot be used as a model since 
its Schwarzchild analogue would be an observer seated 
{\it inside} an expanding star watching its surface expand 
out to infinity. In the case $g_m^{-2} - |X_o|^2 > 0$, 
only the positive sign in Eqn.~(\ref{eqn:amo8}) should be 
chosen for the solution to be physically relevant and it 
is easy to convince oneself that the late time behaviour 
corresponds exactly with that for the case 
$g_m^{-2} - |X_o|^2 = 0$.  When $g_m^{-2} - |X_o|^2 < 0$, 
the minus sign can also be chosen to provide a valid solution 
(choosing the positive sign gives a solution that is 
qualitatively the same as that considered above).  
But, it is easy to check that in this case, the mirror must 
collide with the observer when the late time behaviour 
is considered.  This is not a physically suitable solution.  

\section{Schwarzchild spacetimes: A Classical Calculation}
\label{sec:bh}
We now study the system consisting of a star whose surface is 
silvered and an observer situated well outside it.  
The observer shines monochromatic light of frequency $\omega$ 
on the star and fourier transforms the reflected light. 
We consider two specific star--observer configurations in 
analogy to the mirror--observer ones studied in 
sections~(\ref{sec:accobserver},\ref{sec:mirrorobserver}).  
The first configuration deals with a system in which the star 
is collapsing with the observer being stationary while the 
second deals with a system in which the star is static with 
the observer moving away along a specified trajectory. 
Since these are two distinct cases, they will be dealt with 
in two separate sections. In the first section, the trajectory 
of a radially collapsing surface near the event horizon will 
be derived using the Hamilton-Jacobi formalism for black 
holes~\cite{landau2} and the power spectrum is computed accordingly.  
In the next section, the Schwarzchild analogy of the results in 
section~(\ref{sec:mirrorobserver}) will be derived using the 
Kruskal extension.

\subsection{Thermal spectrum from collapsing star} 
\label{subsec:accstar}

We consider here a collapsing star of mass $M$ whose surface 
moves along an ingoing radial geodesic.  
The spacetime outside the star is just the Schwarzchild geometry.  
The observer is stationed at a constant radius $r_0$  and is 
positioned outside the star for all time.  We work in 
$(1+1)$--dimensions as done in the case of the mirrors. 
It is also assumed that the surface of the star is silvered. 
The  observer shines light along radial geodesics on the 
surface and fourier analyses the reflected light and 
subsequently calculates the power spectrum. 

We will do the following calculations in a more general 
Schwarzchild-like spacetime and then specialise to the 
usual Schwarzchild metric.
Consider the Schwarzchild-like metric in 
$(1+1)$--dimensions.
\begin{equation}
ds^2 = B(r) dt^2 - {dr^2 \over B(r)}  
\label{eqn:gbh1}
\end{equation}
We consider spacetimes that have a horizon at $r = r_h$ 
such that $B(r)$ has the form
\begin{equation}
B(r) \; = \; B_1(r- r_h) \; + \; {\cal O}([r-r_h]^2) 
\label{eqn:horizon}
\end{equation}
near the surface $r = r_h$. $B_1$ is the first derivative 
of $B(r)$ evaluated at $r = r_h$ and is assumed to be non-zero.    
Define a new variable $r^*$,
\begin{equation}
r^* = \int {dr \over B(r)} 
\label{eqn:tortoise}
\end{equation}
to obtain the conformal metric
\begin{equation}
ds^2 = B(r^*)\left[ dt^2 - d(r^*)^2 \right]  
\label{eqn:gbh2}
\end{equation}
The equation of motion of the surface of the star can be 
evaluated by solving the Hamilton-Jacobi  equation in the 
spacetime~\cite{landau2} 
\begin{equation}
g^{ij} (\partial_iS)(\partial_jS) = m^2  
\label{eqn:gbh3}
\end{equation}
where $S$ is Hamilton's characteristic function and $m$ 
is the mass of a particle moving along with the surface of the star.  
Solving for $S$, one has,
\begin{equation}
S = -Et \pm  \int {dr \over B(r)}\sqrt{E^2 - m^2B}  
\label{eqn:gbh7}
\end{equation}
where $E$ is the energy of the particle. 
 The equation of motion of the surface of the star as a 
function of time $t$ is obtained by solving the equation 
\begin{equation}
{\partial S \over \partial E} = {\rm constant} = t_0  
\label{eqn:gbh8}
\end{equation}
Defining $\xi = E/m$, the trajectory of the star's 
surface is,
\begin{equation}
t - t_0 = \pm \xi \int { dr \over B(r)\sqrt{\xi^2 - B} }  
\label{eqn:gbh10}
\end{equation}      
Near the horizon, we use the condition in 
Eqn.~(\ref{eqn:horizon}). Substituting for $B(r)$ into the 
above equation and defining a new variable $x = B_1(r-r_h)$, 
one obtains,
\begin{equation}
t - t_0 = \pm {\xi \over B_1} \int 
{ dx \over x\sqrt{\xi^2 - x} }  
\label{eqn:gbh11}
\end{equation}
Noting that, close to the horizon, the approximation,
\begin{equation}
(\xi^2 - x)^{-1/2} \approx \xi^{-1}\left( 1 + {x \over 2\xi^2} + 
{\cal O}(x^2) \right)  
\label{eqn:gbh12}
\end{equation}
is accurate to ${\cal O}(x^2)$ and choosing the minus 
sign for infalling geodesics, we obtain after performing 
the integration over $x$,
\begin{equation}
t - t_0 = -{1 \over B_1}\left( \ln (x) + {x \over 2\xi^2} + 
{\cal O}(x^2) \right)  \label{eqn:gbh13}
\end{equation}
Substituting for $x$ in terms of $r$, one has,
\begin{equation}
{1 \over B_1}\ln (B_1(r-r_h)) + {r-r_h \over 2\xi^2} + 
{\cal O}\left([r-r_h]^2\right) = -(t-t_0)  
\label{eqn:gbh14}
\end{equation}
Since, near the horizon,
\begin{equation}
r^* \approx \int {dr \over B_1(r-r_h)} = {1 \over B_1}
\ln (B_1(r-r_h)) - {\cal O}(r-r_h) 
\label{eqn:gbh15}
\end{equation}
it is easy to see that, neglecting all powers of $(r-r_h)$ 
greater than the first power, the trajectory assumes the form,
\begin{equation}
r^* + {1 \over 2 B_1\xi^2} e^{B_1 r^*} + {\cal O}
\left(e^{2B_1 r^*}\right) = - (t-t_0)  
\label{eqn:gbh16}
\end{equation}
which, for late times, has the approximate solution
\begin{equation}
r^* = -(t-t_0) - {1 \over 2 B_1\xi^2} e^{-B_1(t-t_0)} - 
{\cal O}\left( e^{-2B_1 (t-t_0)} \right) \label{eqn:gbh17}
\end{equation}
This trajectory, in the conformal co-ordinate $r^*$, 
has exactly the same form as given in Eqn.~(\ref{eqn:latetimes}). 

We specialise henceforth to the usual Schwarzchild spacetime 
with $B(r) = 1 - 2M/r$ which implies $B_1 = 1/2M$. 
The trajectory for the Schwarzchild case is 
\b
r^* \; = \; t_0 - t - {M e^{t_0/2M} \over \xi^2} e^{-t/2M}  
\label{eqn:sbh1}
\e
Since the trajectory given in Eqn.~(\ref{eqn:sbh1}) is 
not physical for all times, we have to set the limits of 
integration from $(t_i, \infty)$ where $t_i>(4M)^{-1}$ 
is chosen so that the above trajectory is a good approximation 
to the actual trajectory.   
Using the results of section~(\ref{sec:accobserver}), the 
power spectrum per unit logarithmic frequency interval is, 
\begin{equation}
{\cal P}(\Omega)\; = \; (4\pi M) \,\left\lbrace{1 \over 2} 
+ N + \sqrt{N(N+1)}\; \cos(2\beta)\right\rbrace,
\label{eqn:sbh2}
\end{equation}
where $N$ is given as usual by Eqn.~(\ref{eqn:planck}) 
but with $\Omega_0$ and $\beta$ given by 
\begin{equation}
\Omega_0 = {1 \over 8\pi M}  \qquad ; \qquad 
\beta = \omega t_0 
\label{eqn:sbh3}
\end{equation}
The inequality that the initial frequency $\omega$ should satisfy 
such that the above spectrum is thermal can be determined 
along the lines of Eqn.~(\ref{eqn:ftrel}), 
\b
\omega \; > \; {\xi^2 \over M}\, e^{-t_0/4M}\, 
e^{(t_i - r_0^*)/4M} 
\label{eqn:sbh4}
\e
where $r_0^*$ is related to $r_0$ by the formula in 
Eqn.~(\ref{eqn:tortoise}). This is the usual ``thermal'' 
power spectrum obtained in the frequency $\Omega$ and is 
analogous to the radiation spectrum obtained in the quantum case.   
In conformal co-ordinates, the solution of the massless 
scalar field are just ordinary plane waves. 
These co-ordinates essentially indicate the obvious point 
that the net redshift caused by the curvature of spacetime 
when light moves from one point to another and back is zero. 
The redshift that is observed is caused solely by reflection 
off the moving surface of the star and that is what gives 
rise to the non-trivial thermal power spectrum.   
\par
If we were to consider a star that collapses for a certain 
time interval $t_i \sim (4M)^{-1}$ to $t_f$, then, by 
applying the results of section~(\ref{sec:accobserver}), 
one obtains a thermal spectrum for frequencies $\Omega > 4M$ 
only when the following inequalities are satisfied.
\begin{equation}
\omega \; > \; {\xi^2 \over M}\, e^{-t_0/4M}\, 
e^{(t_i - r_0^*)/4M} \qquad ; \qquad (t_f - t_i) > 1/4M   
\label{eqn:sbh5}
\end{equation}
Therefore, in this way, the radiation spectrum that arises 
from the collapse of a star to form a black hole has a 
viable classical analogue. 

\subsection{Thermal spectrum from motion of the observer} 
\label{subsec:accobserver}
Here, we attempt to make a correspondence with the results 
of section~(\ref{sec:mirrorobserver}) in the black hole case. 
To do this, we now use the Kruskal co-ordinates in order to 
make a formal correspondence with the Minkowski-Rindler system. 
 Let the Kruskal co-ordinates be denoted by $(T,X)$ and the 
Schwarzchild conformal co-ordinates by $(t,r^*)$. 
The usual trasformations between these two sets of 
co-ordinates are 
\b
T \; = \; e^{r^*/4M} \sinh(t/4M) \qquad ; 
\qquad X \; = \; e^{r^*/4M} \cosh(t/4M) 
\label{eqn:sbo1}
\e
The above transformations are relevant only for the first 
sector of the full Kruskal manifold which represents the 
standard Schwarzchild spacetime outside the event horizon.  
This is the region of interest here. 
\par
Consider a static, spherically symmetric star of mass $M$ 
with a radius $r_0 > 2M$. The star is at rest in the usual 
Schwarzchild spacetime. In the Kruskal co-ordinates, 
however, the motion of the surface of the star is given 
by the equation,
\b
X^2 - T^2 \; = \; e^{r_0^*/2M} 
\label{eqn:sbo2}
\e
Now, consider an observer moving in a ``frame of reference'' 
whose transformations with respect to the Kruskal co-ordinates 
are given by
\b
T \; = \; e^{x^*/4M'} \sinh(\tau/4M') \qquad ; 
\qquad X \; = \; X_o + e^{x^*/4M'} \cosh(\tau /4M') 
\label{eqn:sbo3}
\e
where $(\tau, x^*)$ are the (conformal) co-ordinates in the 
observer`s frame, $X_o>0$ is an arbitrary constant and 
$M' \neq M$ is also a constant. The observer's frame 
of reference, in the $(\tau, x)$ co-ordinates ($x$ is 
related to $x^*$ in the same way as $r$ is related to $r^*$), 
resembles the  Schwarzchild frame in that it appears to 
possess a event horizon located at $x = 2M'$.  
This observer hence has a event horizon corresponding to a 
star of mass $M'$.  We will discuss the motion of the observer 
in the standard Schwarzchild spacetime a little later. 
\par
The trajectory of the silvered star's surface as seen from 
the observer's frame of reference is given by substituting 
Eqn.~(\ref{eqn:sbo3}) into Eqn.~(\ref{eqn:sbo2}) and solving 
the resulting quadratic equation in the variable $e^{x^*/4M'}$. 
One therefore obtains,
\b
e^{x^*_m(t)/4M'} \; = \; - X_o\cosh(\tau /4M') \pm 
\sqrt{X_o^2\cosh^2(\tau /4M') + (e^{r_0^*/2M} - X_o^2) } 
\label{eqn:sbo4}
\e
By construction, the above equation has the same form as 
that in Eqn.~(\ref{eqn:amo4}).  
Hence we can use all the results of that section in 
an identical manner. If $X_o = 0$, it is clear that the 
observer's frame of reference and the standard Schwarzchild 
spacetime are identical and so the star's surface appears to 
be at rest.  This is a trivial solution. Further, to obtain 
a physically viable trajectory for the star, one must 
have $e^{r_0^*/2M} - X_o^2 > 0$.  In this case, it is clearly 
seen that for late times $\tau > 4M'$, the power spectrum seen 
by the observer upon observing the star is of the same 
form as in Eqn.~(\ref{eqn:sbh2}) with $M$ replaced by $M'$ 
and with 
\b
\Omega_0 = {1 \over 8\pi M'} \qquad ; \qquad 
\beta = \omega\ln\left[ { e^{r_0^*/2M} - X_o^2 \over X_o}\right] 
\label{eqn:sbo5}
\e
Thus we see that the spectrum seen by the observer is independent 
of the mass of the star and is dependent only on $M'$.  
Classically, at least, we have a situation where the motion 
of the observer produces the redshift required to obtain 
a thermal power spectrum. 
\par
The motion of the observer (sitting at a constant $x^*_0$ 
in his/her frame) in the standard Schwarzchild spacetime 
can be determined by eliminating the Kruskal co-ordinates 
from Eqn~(\ref{eqn:sbo1}) and  Eqn~(\ref{eqn:sbo3}) to give
\b
e^{r^*/2M} \; = \; e^{x^*_0/2M'} + X_o^2 + 2X_o e^{x_0/4M'} 
\cosh(\tau /4M')   \qquad ; \qquad \tanh(t/4M) \; 
= \; { e^{x_0/4M'} \sinh(\tau /4M') \over X_o + e^{x_0/4M'} 
\cosh(\tau /4M')} \label{eqn:sbo6}
\e
For late times $\tau > 4M'$ which is of interest here, 
the trajectory of the observer in the standard Schwarzchild 
spacetime has the form
\b
r^* \; = \; 2M\ln\left[X_o e^{x_0/4M'}\right] + {M \over 2M'} \, 
\tau + 2M \, e^{-\tau/2M'} \qquad ; \qquad t = {M \over M'} 
\tau  \label{eqn:sbo7}
\e
Thus, we see that if the observer moves along the above 
trajectory, then the power spectrum seen has a thermal nature. 
This result is interesting since it implies the equivalence of 
the motion of the star vis a vis the observer. The phenomenon 
of the collapse of a star to form an event horizon is 
equivalent to a situation where the star is static and the 
observer is moving away from it along a trajectory given above.  
Both situations produce the same thermal spectrum in the 
observer's frame of reference.      

\section{Conclusions}
In conclusion, it is seen that viable classical models that 
mimic the phenomena of black hole radiation can be constructed. 
We have discussed various mirror trajectories and the conditons 
that give rise to a ``thermal'' power spectrum. For the system 
consisting of an inertial mirror viewed by an accelerated observer, 
only situations relevant to the Schwarzchild spacetime are studied. 
The mirror moves along a geodesic in Minkowski spacetime with the 
observer in non-geodesic motion. For a observer moving along an 
accelerated trajectory for a finite interval, a thermal spectrum 
is obtained only if the time interval is large compared to $(1/g)$ 
where $g$ is the acceleration.  Such a result has an analogy  
in the Schwarzchild spacetime in which a star collapses for 
a finite time interval {\it without} forming a black hole.  
The time interval of collapse must be large compared to the 
inverse of the surface gravity of the star which is $(4M)^{-1}$. 
In this case too, the star collapses along a geodesic of the 
spacetime while the observer, who is stationary, is in 
non-geodesic motion. 
\par
The reverse situation in which an uniformly accelerated mirror 
is viewed by an  inertial observer does {\it not} yeild a 
thermal power spectrum (see Appendix~(\ref{sec:accmirror}) 
for details).  This is analogous to the well known quantum 
result in Ref.~\cite{davies77}. In this situation, it the 
observer who is moving along a geodesic with the mirror 
moving along a non-geodesic trajectory. Therefore, it 
appears that the thermality of the power spectrum seems 
to be dependent on whether the mirror or the observer is 
in geodesic motion. 
\par
However, a thermal power spectrum also arises in the case of 
a uniformly accelerated mirror viewed in an uniformly 
accelerated observer's frame for late times.  
In this case, both the mirror and the observer move in 
non-geodesic motion. This situation has a simple analogy 
in the Schwarzchild spacetime in which an observer moving 
along a late time trajectory given Eqn.~(\ref{eqn:sbo7}) 
sees a thermal spectrum when viewing a static 
non-collapsing star. The spectrum seen  is independent of 
the mass $M$ of the star and depends only on the parameters 
defining the obsever's frame of reference. 
We will discuss the quantum analogue of the classical system 
of an accelerated mirror viewed in an accelerated observer's 
frame and its corresponding Schwarzchild system 
in a future publication.  

\section*{\centerline {Acknowledgments}}

\noindent
KS is being supported by the Senior Research Fellowship of the 
Council of Scientific and Industrial Research, India. 

\appendix

\section{Accelerated mirror viewed in Inertial frame} 
\label{sec:accmirror}
We will now consider the reverse case of an inertial 
observer and an accelerating mirror. The trajectory of an 
uniformly accelerating ``receding'' mirror is
\begin{equation}
gx(t) = 1 - \sqrt{1 + g^2 t^2} \label{eqn:acc1}
\end{equation}
where we have assumed the initial conditions that the position 
and velocity are zero at $t=0$.  The reflected wave is easily 
found to be given by the relation
\begin{equation}
\Phi_R(u) = -\exp\left(-i{\omega u \over gu + 1} \right)  
\label{eqn:acc2}
\end{equation}
Assume that the observer is sitting at a distance $x_0$ from 
the origin of co-ordinates.  The FT of the reflected wave is 
\begin{equation}
\tilde{\Phi}_R(\Omega) = -\int_{-\infty}^{\infty}\! dt \, 
e^{-i\Omega t} \exp\left(-i{\omega (t - x_0) \over g(t-x_0) + 1} 
\right)  \label{eqn:acc3}
\end{equation}
where it is assumed that $\Omega > 0$. Making the change 
of variable $y = g(t-x_0) + 1$, we obtain
\begin{eqnarray}
\tilde{\Phi}_R(\Omega) &=& -{e^{(\Omega - \omega)/g} 
\over g}e^{-i\Omega x_0} \int_{-\infty}^{\infty} \! dy\, 
\exp \left(-{i \Omega \over g}x + {i \omega \over g}{1\over x} 
\right) \nonumber \\
&=& -{e^{(\Omega - \omega)/g} \over g}e^{-i\Omega x_0}
\left[ \int_{0}^{\infty} \! dx\, \exp \left(-{i \Omega \over g}x 
+ {i \omega \over g}{1\over x} \right) \, + \, 
\int_{0}^{\infty}\! dx\, \exp \left({i \Omega \over g}x - 
{i \omega \over g}{1\over x} \right) \right] \nonumber \\
&=&  -{e^{(\Omega - \omega)/g} \over g}e^{-i\Omega x_0}
\left[ {\sqrt{2 \omega \Omega} \over i\Omega} + 
{\sqrt{2 \omega \Omega} \over -i\Omega}\right] 
K_1(\sqrt{2\omega\Omega}/g) \equiv 0  \qquad \qquad 
{\rm for} \;\; \Omega > 0 
\label{eqn:acc4}
\end{eqnarray}
where we have used the formula in Ref.~\cite{gandr80}
\begin{equation}
\int_{0}^{\infty} \! dy\, \exp \left(-Ay - {B \over y}\right) 
= {\sqrt{2AB}\over A} K_1(\sqrt{2AB}) 
\label{eqn:formula2}
\end{equation}
with $K_1(x)$ being the modified Bessel 
function of order $1$. 
We see that the FT taken over the entire trajectory is 
identically zero.  If instead of the trajectory given 
above in Eqn.~({\ref{eqn:acc1}), we assume that the mirror 
is inertial to start with and then accelerates continuously, 
we do get a non-zero spectrum.  Assume that the trajectory 
of the mirror is now
\begin{equation}
gx(t) = \left\{ \begin{array}{ll}
                0 & t \leq 0 \\
                1 - \sqrt{1 + g^2 t^2} & t > 0
                \end{array}
        \right. \label{eqn:acc5}
\end{equation}
The reflected wave is now given by
\begin{equation}
\Phi_R(u) = \left\{ \begin{array}{ll}
                    -e^{-i\omega u} & t \leq 0 \\
                     -\exp\left(-i{\omega u \over gu + 1} 
\right) & t > 0
                     \end{array}
            \right. \label{eqn:acc6}
\end{equation}
Assuming as usual that the observer is seated at a 
distance $x_0$ from the origin, the FT of the reflected 
wave for $t>0$ yields a non-zero result.
Therefore, we have, after making the previous change of 
variable,
\begin{eqnarray}
\tilde{\Phi}_R(\Omega) &=& \int_{-\infty}^{\infty}\! dt\, 
e^{-i\Omega t} \Phi_R(t-x_0) \nonumber \\
&=& -e^{i\omega x_0}\int_{0}^{\infty} dt e^{i(\Omega + \omega)t}  
-{e^{(\Omega - \omega)/g} \over g}e^{-i\Omega x_0} 
\int_{0}^{\infty} dx \exp \left(-{i \Omega \over g}x + 
{i \omega \over g}{1\over x} \right) \nonumber \\
&=& -e^{i\omega x_0}{i \over \Omega + \omega}-{e^{i(\Omega - 
\omega)/g}\over g} e^{-i\Omega x_0} {\sqrt{2 \omega \Omega} 
\over i\Omega} K_1(\sqrt{2\omega\Omega}/g) \label{eqn:acc7}
\end{eqnarray}
Taking the modulus square of the above fourier amplitude we 
obtain the required power spectrum. The result is definitly 
not a Planck spectrum of the form given in Eqn.~(\ref{eqn:ps}).  
Thus, a scenario where an accelerated observer looks at an 
inertial detector is not the same as an inertial detector 
looking at an accelerated detector.  It appears to be important 
as to which is moving along an inertial trajectory or in geometric 
terms, which of the two is moving along a geodesic of the spacetime.


\begin{thebibliography}{15}
\bibitem{klt97}
K.~Srinivasan, L.~Sriramkumar and T.~Padmanabhan, `Possible Quantum
Interpretation of Certain Power Spectra in Classical Field Theory',
IJMP~-~D, {\bf 6} 607 (1997).
\bibitem{bandd82}
N.~D.~Birrell and P.~C.~W.~Davies, {\sl Quantum Field Theory in Curved 
Space},\/ Cambridge University Press, Cambridge, pp.~50--54 (1982).
\bibitem{fulling73} 
S.~A.~Fulling, Phys.\ Rev.\ D {\bf 7}, 2850 (1973).
\bibitem{unruh76}
W.~G.~Unruh, Phys.\ Rev.\ D {\bf 14}, 870 (1976).
\bibitem{carlitz87}
Robert D.~Carlitz, Raymond S.~Willey, Phys.\ Rev.\ D {\bf 36} 2327 (1987).
\bibitem{reuter89}
M.~Reuter, C.~T.~Hill, Annals of Physics {\bf 195} 190 (1989).
\bibitem{gerlach88} 
U.~H.~Gerlach, `Heuristic viewpoint concerning the thermal 
ambience relative to an accelerated frame' in {\sl Between Quantum 
and Cosmos},\/ Eds. W.~H.~Zurek  {\it et. al.}, Princeton University 
Press, Princeton, New Jersey (1988).
\bibitem{gerlach76}
U.~H.~Gerlach, Phys.\ Rev.\ D\/ {bf 14}, 1479 (1976).
\bibitem{boyer80}
T.~H.~Boyer, Phys.\ Rev.\ D\/ {\bf 21}, 2137 (1980).
\bibitem{landau2}
L.~D.~Landau and E.~M.~Lifshitz, {\sl The Classical Theory of Fields},
Course of Theoretical Physics, Volume~2, Pergamon Press, New York (1975).
\bibitem{rindler66}
W.~Rindler, Am.\ J.\ Phys.\ {\bf 34}, 1174 (1966).
\bibitem{gandr80}
I.~S.~Gradshteyn and I.~M.~Ryzhik, {\sl Table of Integrals, Series and
Products},\/ Academic, New York (1980). 
\bibitem{davies77}
P.~C.~W.~Davies, S.~A.~Fulling, Proc.\ R.\ Soc.\ Lond.\ A. 
{\bf 356} 237 (1977).
\bibitem{fulling76}
S.~A.~Fulling, P.~C.~W.~Davies, Proc.\ R.\ Soc.\ Lond.\ A. 
{\bf 348} 393 (1976).

\end{thebibliography}
\end{document}